\documentclass{nature_fig}

\usepackage{sansmath}
\usepackage{amssymb}
\newcommand{\mysf}[1]{\mbox{\sansmath$\mathsf{#1}$}}

\usepackage{epsfig}
\usepackage{graphicx}
\usepackage{upgreek}
\usepackage{xcolor}

\title{A progenitor population for bright double neutron star mergers}

\author{R.~D.~Ferdman$^{1*}$, P.~C.~C.~Freire$^2$,
B.~B.~P.~Perera$^3$,
N.~Pol$^{4}$,
F.~Camilo$^5$,
S.~Chatterjee$^6$,
J.~M.~Cordes$^6$
F.~Crawford$^7$,
J.~W.~T.~Hessels$^{8,9}$,
V.~M.~Kaspi$^{10}$,
M.~A.~McLaughlin$^4$,
E.~Parent$^{10}$,
I.~H.~Stairs$^{11}$,
J.~van~Leeuwen$^9$
}

\newcommand{\msun}{M_{\odot}}
\newcommand{\omdot}{\dot{\omega}}
\newcommand{\pbdot}{\dot{P}_b}
\newcommand{\psr}{PSR~J1913+1102}

\newcommand{\apj}{The Astrophysical Journal}

\newcommand{\mnras}{Monthly Notices of the Royal Astronomical Society}
\newcommand{\pasa}{Publications of the Astronomical Society of the Pacific}
\newcommand{\ptrsa}{Philosophical Transactions of the Royal Society A} 

\begin{document}

\maketitle

\begin{affiliations}
 \item Faculty of Science, University of East Anglia, Norwich Research Park, Norwich NR4~7TJ, United Kingdom
 \item Max-Planck-Institut f\"{u}r Radioastronomie, Auf dem H\"{u}gel 69, 53121 Bonn, Germany
 \item Arecibo Observatory, HC3 Box 53995, Arecibo, PR 00612, USA
 \item Department of Physics and Astronomy and Center for Gravitational Waves and Cosmology, West Virginia University, Morgantown, WV 26506-6315, USA
 \item South African Radio Astronomy Observatory, Cape Town, South Africa
 \item Cornell Center for Astrophysics and Planetary Science and Department of Astronomy, Cornell University, Ithaca, NY 14853, USA
 \item Department of Physics and Astronomy, Franklin and Marshall College, Lancaster, PA 17604-3003, USA
 \item Anton Pannekoek Institute for Astronomy, University of Amsterdam, Science Park 904, 1098 XH, Amsterdam, The Netherlands
 \item ASTRON, Netherlands Institute for Radio Astronomy, Oude Hoogeveensedijk 4, 7991 PD Dwingeloo, The Netherlands
 \item Department of Physics and McGill Space Institute, McGill University, Montreal, QC H3A~2T8, Canada
 \item Department of Physics and Astronomy, University of British Columbia, 6224 Agricultural Road, Vancouver, BC V6T~1Z1, Canada
 \item[*] email: r.ferdman@uea.ac.uk
\end{affiliations}

\begin{abstract}
The discovery of a radioactively powered kilonova associated with the binary neutron star merger GW170817 
was the first---and still only---confirmed electromagnetic counterpart to a gravitational-wave event\cite{ligo17a,ligo17b}. However, observations of late-time electromagnetic emission are in tension with the expectations from standard neutron-star merger models. Although the large measured ejecta mass\cite{ligo17c, cbv+17} is potentially explained by a progenitor system that is asymmetric in terms of the stellar component masses, i.e.~with a mass ratio $q$ of 0.7--0.8\cite{pan18}, 
the known Galactic population of merging double neutron star (DNS) systems (i.e.~those that will coalesce within billions of years or less) has, until now, only consisted of nearly equal-mass ($q > 0.9$) binaries\cite{tkf+17}.
PSR~J1913+1102 is a DNS system 
in a 5-hour, low-eccentricity ($e = 0.09$) orbit,
implying an orbital separation of 1.8 solar radii\cite{lfa+16}, with the two neutron stars predicted to coalesce in 470 million years due to gravitational-wave emission.
Here we report that the masses of the two neutron stars, as measured by a dedicated pulsar timing campaign, are $1.62 \pm  0.03$ and $1.27 \pm  0.03$ solar masses for the pulsar and companion neutron star, respectively; 
with a measured mass ratio $q = 0.78 \pm 0.03$, it is the most asymmetric DNS among known merging systems. 
Based on this detection, our population synthesis analysis implies that such asymmetric binaries represent between 2 and 30\% (90\% confidence) of the total population of merging DNS binaries. The coalescence of a member of this population offers a possible explanation for the anomalous properties of GW170817, including the observed kilonova emission from that event. 
\end{abstract}

We have been regularly monitoring the DNS \psr{} with the Arecibo radio telescope since its discovery\cite{lfa+16} in 2012. 
Our observations have used the Puerto Rico Ultimate Pulsar Processing Instrument (PUPPI) to coherently remove dispersive smearing from the pulsar signal,
caused by the interstellar free-electron plasma along the line of sight to the pulsar. We analysed data from this pulsar using standard pulse timing techniques (see Methods).

With a spin period of 27 ms, \psr{} was the first-formed neutron star (NS) in this 
binary system; this was subsequently recycled by accretion of matter from the
progenitor to the second NS \cite{lfa+16}. Its timing has allowed a precise measurement of the rate of advance of periastron, $\omdot= 5.6501\pm 0.0007$ degrees per year. Assuming general relativity (GR), this leads to a total system mass measurement of $2.8887 \pm 0.0006\,\msun$---the most massive among known DNS systems (by a $2\%$ margin). 
In addition, we have now determined two more post-Keplerian (PK) parameters: the first, the Einstein delay ($\gamma = 0.471 \pm 0.015\,$ms), describes the effect of gravitational redshift and relativistic time dilation, due to the varying orbital velocity and proximity of the neutron stars to one another during their orbits. 
The second is the variation of the orbital period caused by the emission of gravitational waves ($\pbdot = (-4.8 \pm 0.3) \times 10^{-13}\,$s\,s$^{-1}$).

\begin{figure}
\includegraphics[width=\textwidth]{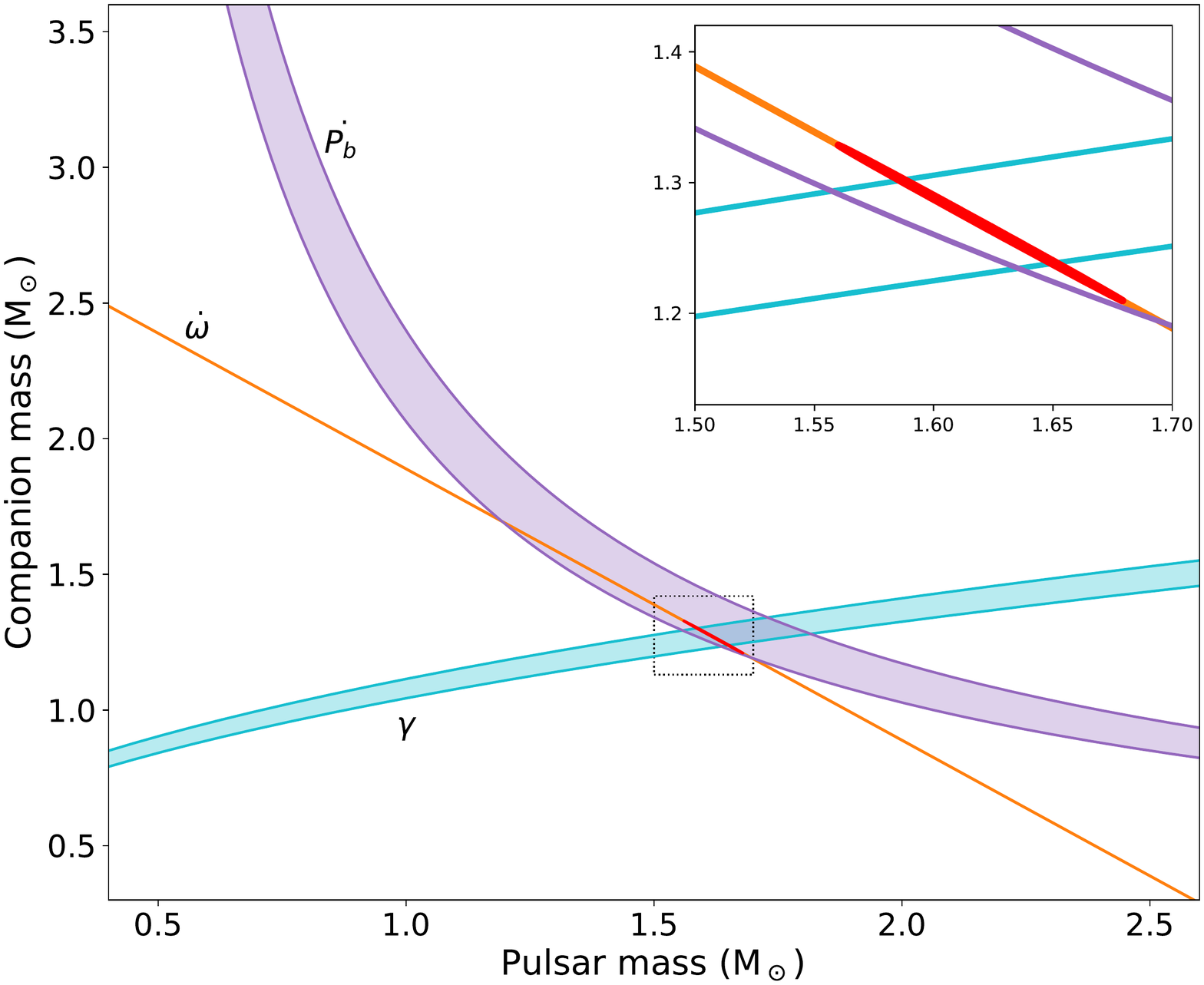} 
\caption{Pulsar mass/companion mass diagram for the PSR~J1913+1102 system.  
Shaded regions bounded by solid curves represent $1\sigma$ mass constraints from each measured post-Keplerian parameter, derived in the context of general relativity. These are: orbital precession rate ($\omdot$), time dilation/gravitational redshift ($\gamma$), and the rate of orbital decay ($\pbdot$).  The inset shows a zoom-in of the dotted square region in the main plot, with the $3\sigma$ confidence region for the mass measurements shaded in red.
The two most precisely measured parameters allow us to determine the individual masses of this system; each additional post-Keplerian parameter measurement provides an independent consistency test of the predictions of general relativity.\label{fig:pkparams}}
\end{figure}

In Figure~\ref{fig:pkparams}, we show the GR mass constraints corresponding to each measured PK parameter.
Combining $\omdot$ and $\gamma$ we obtain the individual NS masses:  $m_p = 1.62 \pm 0.03\,\msun$ and $m_c = 1.27\pm 0.03\,\msun$ for the pulsar and the companion respectively, the mass ratio is $q = m_c/m_p = 0.78 \pm 0.03$. The observed $\pbdot$ is consistent with the GR prediction for these NS masses; apart from confirming them, this effect has already provided a unique test of alternative gravitational theories that will be reported elsewhere\cite{freire19}.
Table~\ref{tab:params} summarises the best-fit model parameters for the \psr{} system.

\psr{} is part of a population of several very close DNS binary systems with moderate orbital eccentricities ($\lesssim 0.2$) and low proper motions (e.g.~PSRs~J0737$-$3039A/B\cite{ksm+06} and J1756$-$2251\cite{fsk+14}).  These imply an evolutionary path in which the second-formed NS was born as a result of an envelope-stripped helium star progenitor having undergone a supernova with very little mass loss and low natal kick\cite{tlm+13, tkf+17}, or a rapid core-collapse event due to electron capture onto an oxygen-neon-magnesium core\cite{miyaji80,nomoto84,podsi05}. Either of these scenarios lead to a low-mass NS\cite{lfa+16}, which is confirmed by our measurements.

The \psr{} mass ratio makes it the most asymmetric among the known DNS binaries that are expected to merge within a Hubble time, which otherwise have $q\gtrsim 0.9$.
Considering all known DNS systems, the only one having a similar mass asymmetry is PSR~J0453+1559 \cite{msf+15}, with $q=0.753\pm 0.005$;  
however, its orbital period of 4.07 days implies a coalescence time $\sim\! 100$ times greater than the age of the Universe.
In contrast, \psr{} has an expected time to coalescence of 470\,Myr, which we determined from  the orbital decay rate and other measured orbital elements.

\begin{figure}
\begin{center}
\includegraphics[width=\textwidth]{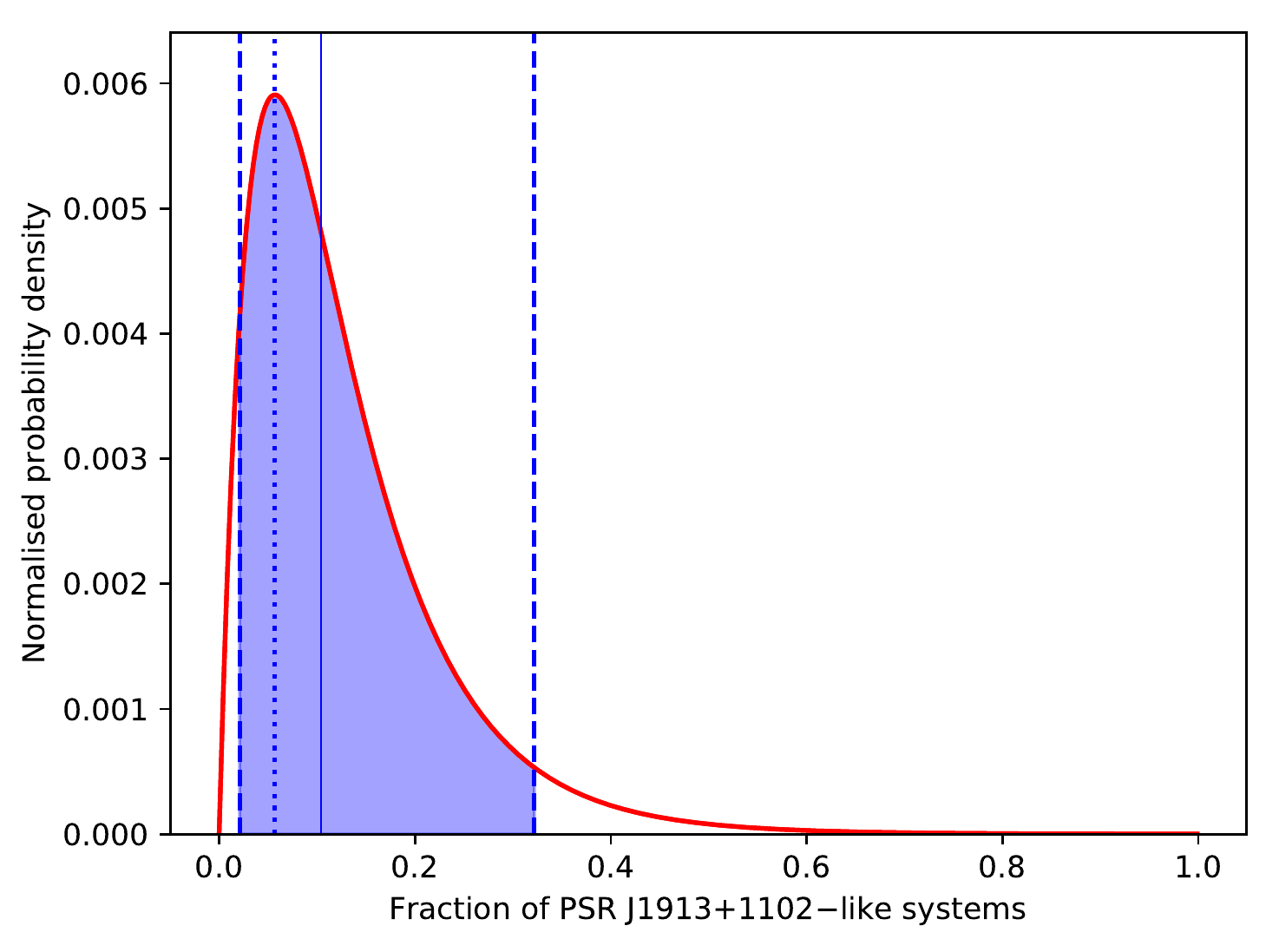}
\end{center}
\caption{
Probability density of the population of \psr{}-like DNS systems in the Galaxy, as a fraction of the total number of DNSs that will merge within a Hubble time.
We find this fraction to be $0.11^{+0.21}_{-0.09}$, where the uncertainty represents the 90\% confidence interval (represented by vertical dashed lines). The quoted value is the median of the distribution, shown on the plot as a solid vertical line, and the peak value of 0.06 represented by a dotted vertical line.  This implies that roughly 1 in 10 merging DNS systems are likely to be asymmetric in component masses. 
\label{fig:pop_frac}}
\end{figure}

There are currently nine confirmed compact DNS binaries that are predicted to merge within a Hubble time, for which precise NS mass measurements have been made\cite{tkf+17}.  
We have performed a population synthesis analysis for these DNS systems, including \psr{}, using their individual properties and known masses (see also Methods). We find that \psr{}-like binaries represent $11_{-9}^{+21}\,\%$ of merging DNS systems (Figure~\ref{fig:pop_frac}), where the quoted value is the median and the errors represent the 90\% confidence intervals.
This therefore establishes the existence of a population of asymmetric DNS systems that is significant enough to potentially lead to the discovery of several corresponding merger events by ground-based gravitational-wave observatories such as LIGO/Virgo.  As such, the discovery of \psr{} provides evidence for the need to account for asymmetric coalescing DNS binaries---approximately one tenth of events---in understanding merger scenarios and the physics that underpins them.


Observations of the electromagnetic counterparts to the GW170817 event have largely been related to a relatively large amount of ejecta due to the preceding DNS merger, with mass on the order of 0.05\,$\msun$\cite{cbv+17, kmb+17}. This is in tension with standard models of DNS coalescence, which typically predict at least a factor of 5 smaller ejecta mass, primarily based on the assumption of equal-mass (or near-equal mass) progenitor DNS binary systems\cite{hkk+13,pan18,rd19}.  Until recently, this has been a reasonable assumption given the known DNS population\cite{tkf+17}. It is plausible that the anomalously massive ejecta inferred from the observed late-time emission, in the specific case of GW170817, may be explained with an equal-mass system,
particularly if one includes a secular component to the ejecta\cite{Siegel18, Fernandez18, rph+18}. 
The merger may also be explained by the invocation of various models to describe the observations. These include an off-axis jet from a short gamma-ray burst\cite{tpv+17, hnr+17}; a mildly relativistic wide-angle outflow that interacts with the dynamic ejecta\cite{mnh+18, pk18}; a hypermassive neutron star remnant of the merger acting as a spin-down energy source\cite{mtq18, llyz18}; as well as a hierarchical triple system in which a disk is formed from a Roche lobe-filling outer star\cite{cm18}.

In contrast, numerical simulations have shown that high-asymmetry ($0.65\lesssim q \lesssim 0.85$) systems will naturally produce larger tidal distortions during the merger phase, and result in larger-mass and therefore brighter disks than for roughly equal-mass systems\cite{st06, rbg+10, du17, rd19}. The resulting tidal effects consistently produce sufficient neutron-rich ejecta to power a kilonova and result in an enhancement of r-process material\cite{llp+16}.
A sufficiently unequal-mass binary that will merge within a Hubble time may therefore be responsible for events such as GW170817,
particularly in the slow, $\sim\! 0.04\,\msun$ red component seen in the latter\cite{cbv+17}.  
A significant population of asymmetric DNSs such as \psr{}
would therefore lead to an enhanced detection rate of 
bright kilonovae.  The electromagnetic counterparts to such events are therefore particularly important for understanding the Galactic heavy-element abundance\cite{tlf+13,jbp+15,kmb+17}. 

Enhanced pre-coalescence tidal distortions due to asymmetric mergers may also allow for certain NS equation of state models to be ruled out through study of the gravitational-wave waveform\cite{rms+09, lw15, amd+15,ligo18, rph+18}. 
Additionally, gravitational waves from merging DNSs have recently been used as distance indicators (so-called ``standard sirens''), allowing for an independent probe of the Hubble constant, $H_0$, when combined with radial velocity measurements of the electromagnetic counterparts\cite{ligo17d}. 
Future asymmetric DNS mergers with similar electromagnetic counterparts to those of GW170817 would lead to a significantly more precise determination of $H_0$---an estimated $\sim 15$ suitable detections would provide a $\sim 2\%$ measurement\cite{hng+18}---and potentially provide the means by which to resolve the tension that exists between $H_0$ as measured with the cosmic microwave background \cite{planck18} and through local Universe analysis methods\cite{rcy+19}.



\begin{figure}
\begin{center}
\includegraphics[width=\textwidth]{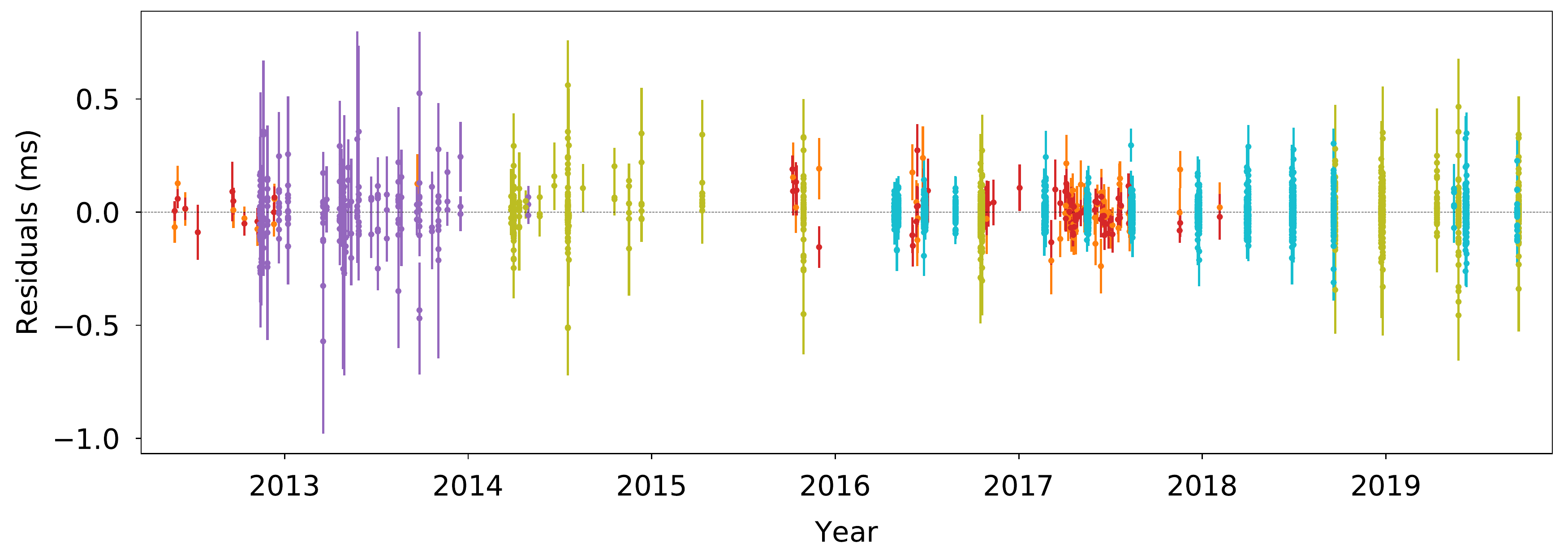}
\end{center}
\caption{
Post-fit timing residuals for \psr{}.  These are obtained after including all best-fit parameters in the DDGR model ephemeris for this pulsar. Each contributing instrument is represented by different colours, as follows: Mock spectrometer centred at 1300\,MHz and 1450\,MHz---orange and red, respectively; PUPPI centred at 1400\,MHz in incoherent mode---purple; PUPPI centred at 1400\,MHz in coherent folding mode---yellow; and PUPPI centred at 2350\,MHz in coherent fold mode---cyan.  The latter provided a significant improvement in data quality, as evidenced by the reduction in weighted rms timing residual to $48\,\mu$s, down from $72\,\mu$s at L-band coherent mode. 
\color{red}{Error bars shown reflect the $1\sigma\ (68\%)$ uncertainties of each data point.}
\label{fig:J1913_resids}}
\end{figure}

\begin{methods}

\subsection{Timing analysis.}
\psr{} was discovered in 2012 by Einstein{@}Home in data from the PALFA survey\cite{lfa+16}, which uses the William E.~Gordon 305-m radio telescope at the Arecibo Observatory (AO) in Puerto Rico, to search for pulsars within 5 degrees of the Galactic plane. Since the discovery, we have been using AO to regularly monitor this pulsar with the Mock Spectrometer and the Puerto Rico Ultimate Pulsar Processing Instrument (PUPPI) pulsar backend system. This has been done as both a dedicated follow-up campaign of \psr{} and regularly as a test source before PALFA observing sessions.  

The times of arrival of these pulses were measured by cross-correlating each pulse profile with a noise-free representative template profile, from which we calculate a phase shift and apply it to the observed time stamp of the data profile\cite{tay92}.  The uncertainty in each resulting pulse arrival time is determined by adopting the error in the calculated phase shift from the aforementioned correlation procedure. This process was carried out using the \textsc{PSRCHIVE} suite of analysis tools\cite{psrchive04}.
Corrections between Terrestrial Time and the observatory clock were applied using data from the Global Positioning System (GPS) satellites and the Bureau International des Poids et Mesures. 
Our model also included input from the Jet Propulsion Laboratory DE436 solar system ephemeris, in order to convert measured arrival times to the reference frame of the Solar System barycentre, by taking into account the motion of the Earth.

These barycentred times of arrival were then compared to a predictive model of their expected arrival at Earth using the \textsc{TEMPO}\cite{tempourl}
pulsar timing software package.  
Every rotation of the NS is enumerated relative to a reference observing epoch,
by accounting in our model for intrinsic pulsar properties such as the rotation frequency and its spin-down rate, as well as its sky position and proper motion.  We have also addressed potential arrival-time delays due to the frequency-dependent refractive effect of the ionised interstellar medium, by including the dispersion measure (DM) in our timing model, which is the integrated column density of free electrons along the line of sight between Earth and the pulsar.  
The relatively large value $\mathrm{DM}=339.026\pm0.005$\,pc\,cm$^{-3}$ for this pulsar explains why initial observation, primarily taken at an observing frequency centred at 1.4 GHz, displayed evidence of interstellar scattering that resulted in significant smearing of the observed pulse shape due to multi-path propagation of the signal on its way to Earth\cite{lfa+16}. This led to increased systematic uncertainties in the derived pulsar parameters, and we therefore switched to observations with the higher-frequency S-band-low receiver (centred at 2.4 GHz with a bandwidth of 800\,MHz) to reduce these effects.

Along with these model parameters, our timing data resulted in significant measurement of the Keplerian orbital elements of the \psr{} system, as well as several post-Keplerian parameters; the latter are a theory-independent set of parameters that characterise perturbations on the Keplerian description of the orbit in the relativistic regime\cite{dd85,dd86,dt92}. As described in the main text of this letter, we have increased the measurement precision of the orbital precession rate and have determined the Einstein delay, from which we have been able to constrain the individual masses of the NSs in this system. We now have also made a precise determination of the orbital decay rate, due to the emission of gravitational waves, which serve to remove orbital energy from the system over time.  We expect the precision of the orbital decay to improve rapidly over time with further observations.   It should also be noted that sources of kinematic biases can be introduced into the measured orbital decay and pulsar spin-down rates, from apparent acceleration of the pulsar due to its tangential motion (i.e.~the ``Shklovskii effect''\cite{shk70}) and the Galactic potential\cite{nt95,pbm+19}. We find the total proper motion of this pulsar to be $9.3 \pm 0.9$\,mas\,yr$^{-1}$, within $3\sigma$ of what would be expected if the \psr{} system were in the Local Standard of Rest ($6.50$\,mas\,yr$^{-1}$); this assumes a distance of 7.14\,kpc, based on the measured value of DM, and calculated using a model of the Galactic ionised electron density distribution\cite{ymw17}.  The total kinematic bias to the observed orbital decay corresponds to approximately one-third of the uncertainty in the orbital decay measurement; we are therefore confident that our measurements are consistent with intrinsic parameter values for the pulsar, at the current level of uncertainty.

Once we apply our model to the data set, we produce post-fit timing residuals---the difference between the predicted and observed pulse arrival times (Figure~\ref{fig:J1913_resids}).
The timing precision achieved by this fit to our pulse arrival-time data is characterised by the root-mean-square (rms) of the post-fit timing residuals.  Our analysis of the \psr\ data set resulted in rms residuals of $56.1\,\mu$s, consistent with the typical measured uncertainty in the observed pulse arrival times. We achieved a reduced $\chi^2$ (i.e.~$\chi^2$ divided by the number of degrees of freedom) of 1.01 for our fit, reaffirming the success of our timing model in describing the system, and  implying that the timing residuals can be well represented by white Gaussian noise, as can be seen in Figure~\ref{fig:J1913_resids}.

\subsection{Population synthesis.}
Modelling of the merger event that caused GW170817 has mostly relied on a DNS population consisting of roughly equal-mass neutron stars.  Although it may be the case that it was the result of a binary system with pre-merger mass ratio $q\sim 1$, the discovery of \psr\ highlights the need to consider the effects of an asymmetric DNS merger.
We note here for completeness that the possibility exists for GW170817 to have been cause by a neutron star-black hole merger. However, the abnormally low mass of the black hole in such a progenitor system would make this an unlikely scenario.

Previous studies\cite{kkl03,kpm15,pml19,pml20} simulated the population of a DNS system from its measured parameters, within a modelled Galactic pulsar population, and the sensitivities of the pulsar survey in which it was discovered. The modelling must account for  selection effects, including the search degradation factor due to orbital acceleration, calculated from a semi-analytical model with the pulsar and companion masses and inclination as input\cite{2013MNRAS.432.1303B}. 
We calculated
the probability density of the population of \psr{}-like DNSs that are beamed toward Earth ($N_\mathrm{obs, J1913}$) using the more precisely measured orbital properties presented in this work. 
Assuming a  beaming correction fraction for the pulsar\cite{pml19} of $f_b = 4.6$, we derived the probability density of the total population ($N_\mathrm{tot, J1913}$) of J1913$+$1102-like DNS systems in the Galaxy: ($N_\mathrm{pop, J1913}$ = $N_\mathrm{obs, J1913} \times f_b$). 
The mode of the resulting distribution is    
$N_\mathrm{tot, J1913} = 700^{+2600}_{-400}$, where the uncertainties represent the 90\% confidence interval of the distribution. This is  consistent with previous estimates\cite{pml19}, but has smaller error bars due to the updated orbital parameters and the addition of a new radio pulsar survey\cite{pml20}.

Due to its small orbital period, the \psr{} system will merge within a Hubble time; there are eight other known DNS systems in the Galaxy that will also merge within the age of the Universe (which we henceforth refer to as ``Merging DNSs'', MDNSs). We obtained the individual probability densities of the population of these MDNSs using results given in the aforementioned previous studies\cite{pml19,pml20}. Assuming that these individual population distributions represent independent continuous random variables, we estimate the total population of MDNS systems in the Galaxy by convolving the individual population probability distributions, resulting in a mode $N_\mathrm{tot, MDNS} = (11.4^{+6.3}_{-3.8})\times 10^3$.
Using our derived probability densities of \psr{}-like systems together with those of all MDNS systems, we then compute the probability density of \psr{}-like DNS systems in the Galaxy \textit{as a fraction of the MDNS population} to be $11_{-9}^{+21}\,\%$ (90\% confidence), using the median as the quoted value (with the mode of the distribution occurring at 6\%); Figure~\ref{fig:pop_frac} presents the corresponding probability distribution.  This in turn leads to an estimate that roughly one tenth of detected DNS mergers result from the coalescence of an asymmetric binary system.



\end{methods}


\clearpage

\clearpage

\clearpage

\clearpage


\begin{thebibliography}{10}
\expandafter\ifx\csname url\endcsname\relax
  \def\url#1{\texttt{#1}}\fi
\expandafter\ifx\csname urlprefix\endcsname\relax\def\urlprefix{URL }\fi
\providecommand{\bibinfo}[2]{#2}
\providecommand{\eprint}[2][]{\url{#2}}

\bibitem{ligo17a}
\bibinfo{author}{Abbott, B.} \emph{et~al.}
\newblock \bibinfo{title}{{GW170817:} observation of gravitational waves from a
  binary neutron star inspiral}.
\newblock \emph{\bibinfo{journal}{Physical Review Letters}}
  \bibinfo{pages}{161101} (\bibinfo{year}{2017}).

\bibitem{ligo17b}
\bibinfo{author}{Abbott, B.~P.} \emph{et~al.}
\newblock \bibinfo{title}{Multi-messenger observations of a binary neutron star
  merger}.
\newblock \emph{\bibinfo{journal}{The Astrophysical Journal}}
  \textbf{\bibinfo{volume}{848}}, \bibinfo{pages}{L12} (\bibinfo{year}{2017}).

\bibitem{ligo17c}
\bibinfo{author}{Abbott, B.~P.} \emph{et~al.}
\newblock \bibinfo{title}{Estimating the contribution of dynamical ejecta in
  the kilonova associated with {GW}170817}.
\newblock \emph{\bibinfo{journal}{The Astrophysical Journal}}
  \textbf{\bibinfo{volume}{850}}, \bibinfo{pages}{L39} (\bibinfo{year}{2017}).

\bibitem{cbv+17}
\bibinfo{author}{Cowperthwaite, P.~S.} \emph{et~al.}
\newblock \bibinfo{title}{The electromagnetic counterpart of the binary neutron
  star merger {LIGO}/{Virgo} {GW}170817. {II}. {UV}, optical, and near-infrared
  light curves and comparison to kilonova models}.
\newblock \emph{\bibinfo{journal}{The Astrophysical Journal}}
  \textbf{\bibinfo{volume}{848}}, \bibinfo{pages}{L17} (\bibinfo{year}{2017}).

\bibitem{pan18}
\bibinfo{author}{Pankow, C.}
\newblock \bibinfo{title}{On {GW170817} and the {Galactic} binary neutron star
  population}.
\newblock \emph{\bibinfo{journal}{The Astrophysical Journal}}
  \textbf{\bibinfo{volume}{866}}, \bibinfo{pages}{60} (\bibinfo{year}{2018}).

\bibitem{tkf+17}
\bibinfo{author}{Tauris, T.~M.} \emph{et~al.}
\newblock \bibinfo{title}{Formation of double neutron star systems}.
\newblock \emph{\bibinfo{journal}{The Astrophysical Journal}}
  \textbf{\bibinfo{volume}{846}}, \bibinfo{pages}{170} (\bibinfo{year}{2017}).

\bibitem{lfa+16}
\bibinfo{author}{Lazarus, P.} \emph{et~al.}
\newblock \bibinfo{title}{{Einstein}@{Home} {discovery} {of} a {double}
  {neutron} {star} {binary} {in} {the} {PALFA} {survey}}.
\newblock \emph{\bibinfo{journal}{The Astrophysical Journal}}
  \textbf{\bibinfo{volume}{831}}, \bibinfo{pages}{150} (\bibinfo{year}{2016}).

\bibitem{freire19}
\bibinfo{author}{Freire, P. C.~C.} \emph{et~al.}
\newblock \bibinfo{title}{{Closing the mass gap of spontaneous scalarization
  with a pulsar in an asymmetric double-neutron-star system}}.
\newblock \emph{\bibinfo{journal}{In preparation}}
  (\bibinfo{year}{2020}).

\bibitem{ksm+06}
\bibinfo{author}{Kramer, M.} \emph{et~al.}
\newblock \bibinfo{title}{Tests of general relativity from timing the double
  pulsar}.
\newblock \emph{\bibinfo{journal}{Science}} \textbf{\bibinfo{volume}{314}},
  \bibinfo{pages}{97--102} (\bibinfo{year}{2006}).

\bibitem{fsk+14}
\bibinfo{author}{Ferdman, R.~D.} \emph{et~al.}
\newblock \bibinfo{title}{{PSR J1756$-$2251: a pulsar with a low-mass neutron
  star companion}}.
\newblock \emph{\bibinfo{journal}{Monthly Notices of the Royal Astronomical
  Society}} \textbf{\bibinfo{volume}{443}}, \bibinfo{pages}{2183--2196}
  (\bibinfo{year}{2014}).

\bibitem{tlm+13}
\bibinfo{author}{Tauris, T.~M.} \emph{et~al.}
\newblock \bibinfo{title}{Ultra-stripped type {Ic} supernovae from close binary
  evolution}.
\newblock \emph{\bibinfo{journal}{The Astrophysical Journal}}
  \bibinfo{pages}{L23} (\bibinfo{year}{2013}).

\bibitem{miyaji80}
\bibinfo{author}{{Miyaji}, S.}, \bibinfo{author}{{Nomoto}, K.},
  \bibinfo{author}{{Yokoi}, K.} \& \bibinfo{author}{{Sugimoto}, D.}
\newblock \bibinfo{title}{{Supernova triggered by electron captures.}}
\newblock \emph{\bibinfo{journal}{Publication of the Astronomy Society of
  Japan}} \textbf{\bibinfo{volume}{32}}, \bibinfo{pages}{303--329}
  (\bibinfo{year}{1980}).

\bibitem{nomoto84}
\bibinfo{author}{{Nomoto}, K.}
\newblock \bibinfo{title}{{Evolution of 8-10 solar mass stars toward electron
  capture supernovae. I - Formation of electron-degenerate O + NE + MG cores.}}
\newblock \emph{\bibinfo{journal}{\apj}} \textbf{\bibinfo{volume}{277}},
  \bibinfo{pages}{791--805} (\bibinfo{year}{1984}).

\bibitem{podsi05}
\bibinfo{author}{{Podsiadlowski}, P.} \emph{et~al.}
\newblock \bibinfo{title}{{The double pulsar J0737-3039: testing the neutron
  star equation of state}}.
\newblock \emph{\bibinfo{journal}{\mnras}} \textbf{\bibinfo{volume}{361}},
  \bibinfo{pages}{1243--1249} (\bibinfo{year}{2005}).
\newblock \eprint{astro-ph/0506566}.

\bibitem{msf+15}
\bibinfo{author}{Martinez, J.~G.} \emph{et~al.}
\newblock \bibinfo{title}{{Pulsar J0453+1559: a double neutron star system with
  a large mass asymmetry}}.
\newblock \emph{\bibinfo{journal}{The Astrophysical Journal}}
  \textbf{\bibinfo{volume}{812}}, \bibinfo{pages}{143} (\bibinfo{year}{2015}).

\bibitem{kmb+17}
\bibinfo{author}{Kasen, D.}, \bibinfo{author}{Metzger, B.},
  \bibinfo{author}{Barnes, J.}, \bibinfo{author}{Quataert, E.} \&
  \bibinfo{author}{Ramirez-Ruiz, E.}
\newblock \bibinfo{title}{{Origin of the heavy elements in binary neutron-star
  mergers from a gravitational-wave event}}.
\newblock \emph{\bibinfo{journal}{Nature}} \textbf{\bibinfo{volume}{551}},
  \bibinfo{pages}{80--84} (\bibinfo{year}{2017}).

\bibitem{hkk+13}
\bibinfo{author}{Hotokezaka, K.} \emph{et~al.}
\newblock \bibinfo{title}{{Mass ejection from the merger of binary neutron
  stars}}.
\newblock \emph{\bibinfo{journal}{Physical Review D - Particles, Fields,
  Gravitation and Cosmology}} \textbf{\bibinfo{volume}{87}},
  \bibinfo{pages}{024001} (\bibinfo{year}{2013}).

\bibitem{rd19}
\bibinfo{author}{Radice, D.} \& \bibinfo{author}{Dai, L.}
\newblock \bibinfo{title}{{Multimessenger parameter estimation of GW170817}}.
\newblock \emph{\bibinfo{journal}{European Physical Journal A}}
  \textbf{\bibinfo{volume}{55}}, \bibinfo{pages}{50} (\bibinfo{year}{2019}).

\bibitem{Siegel18}
\bibinfo{author}{{Siegel}, D.~M.} \& \bibinfo{author}{{Metzger}, B.~D.}
\newblock \bibinfo{title}{{Three-dimensional GRMHD Simulations of
  Neutrino-cooled Accretion Disks from Neutron Star Mergers}}.
\newblock \emph{\bibinfo{journal}{\apj}} \textbf{\bibinfo{volume}{858}},
  \bibinfo{pages}{52} (\bibinfo{year}{2018}).
\newblock \eprint{1711.00868}.

\bibitem{Fernandez18}
\bibinfo{author}{Fernández, R.}, \bibinfo{author}{Tchekhovskoy, A.},
  \bibinfo{author}{Quataert, E.}, \bibinfo{author}{Foucart, F.} \&
  \bibinfo{author}{Kasen, D.}
\newblock \bibinfo{title}{{Long-term GRMHD simulations of neutron star merger
  accretion discs: implications for electromagnetic counterparts}}.
\newblock \emph{\bibinfo{journal}{Monthly Notices of the Royal Astronomical
  Society}} \textbf{\bibinfo{volume}{482}}, \bibinfo{pages}{3373--3393}
  (\bibinfo{year}{2018}).
\newblock \urlprefix\url{https://doi.org/10.1093/mnras/sty2932}.
\newblock
  \eprint{https://academic.oup.com/mnras/article-pdf/482/3/3373/26660289/sty2932.pdf}.

\bibitem{rph+18}
\bibinfo{author}{Radice, D.} \emph{et~al.}
\newblock \bibinfo{title}{Binary neutron star mergers: Mass ejection,
  electromagnetic counterparts and nucleosynthesis}.
\newblock \emph{\bibinfo{journal}{The Astrophysical Journal}}
  \textbf{\bibinfo{volume}{869}}, \bibinfo{pages}{130} (\bibinfo{year}{2018}).

\bibitem{tpv+17}
\bibinfo{author}{Troja, E.} \emph{et~al.}
\newblock \bibinfo{title}{{The X-ray counterpart to the gravitational-wave
  event GW170817}}.
\newblock \emph{\bibinfo{journal}{Nature}} \textbf{\bibinfo{volume}{551}},
  \bibinfo{pages}{71--74} (\bibinfo{year}{2017}).

\bibitem{hnr+17}
\bibinfo{author}{Haggard, D.} \emph{et~al.}
\newblock \bibinfo{title}{A deep {Chandra} {X}-ray study of neutron star
  coalescence {GW170817}}.
\newblock \emph{\bibinfo{journal}{The Astrophysical Journal}}
  \textbf{\bibinfo{volume}{848}}, \bibinfo{pages}{L25} (\bibinfo{year}{2017}).

\bibitem{mnh+18}
\bibinfo{author}{Mooley, K.~P.} \emph{et~al.}
\newblock \bibinfo{title}{{A mildly relativistic wide-angle outflow in the
  neutron-star merger event GW170817}}.
\newblock \emph{\bibinfo{journal}{Nature}} \textbf{\bibinfo{volume}{554}},
  \bibinfo{pages}{207--210} (\bibinfo{year}{2018}).

\bibitem{pk18}
\bibinfo{author}{Piro, A.~L.} \& \bibinfo{author}{Kollmeier, J.~A.}
\newblock \bibinfo{title}{Evidence for cocoon emission from the early light
  curve of {SSS17a}}.
\newblock \emph{\bibinfo{journal}{The Astrophysical Journal}}
  \textbf{\bibinfo{volume}{855}}, \bibinfo{pages}{103} (\bibinfo{year}{2018}).

\bibitem{mtq18}
\bibinfo{author}{Metzger, B.~D.}, \bibinfo{author}{Thompson, T.~A.} \&
  \bibinfo{author}{Quataert, E.}
\newblock \bibinfo{title}{A magnetar origin for the kilonova ejecta in
  {GW170817}}.
\newblock \emph{\bibinfo{journal}{The Astrophysical Journal}}
  \textbf{\bibinfo{volume}{856}} (\bibinfo{year}{2018}).

\bibitem{llyz18}
\bibinfo{author}{Li, S.-Z.}, \bibinfo{author}{Liu, L.-D.}, \bibinfo{author}{Yu,
  Y.-W.} \& \bibinfo{author}{Zhang, B.}
\newblock \bibinfo{title}{What powered the optical transient {AT2017gfo}
  associated with {GW170817?}}
\newblock \emph{\bibinfo{journal}{The Astrophysical Journal}}
  \textbf{\bibinfo{volume}{861}}, \bibinfo{pages}{L12} (\bibinfo{year}{2018}).

\bibitem{cm18}
\bibinfo{author}{Chang, P.} \& \bibinfo{author}{Murray, N.}
\newblock \bibinfo{title}{{GW170817: A neutron star merger in a
  mass-transferring triple system}}.
\newblock \emph{\bibinfo{journal}{Monthly Notices of the Royal Astronomical
  Society: Letters}} \textbf{\bibinfo{volume}{474}}, \bibinfo{pages}{L12--L16}
  (\bibinfo{year}{2018}).

\bibitem{st06}
\bibinfo{author}{Shibata, M.} \& \bibinfo{author}{Taniguchi, K.}
\newblock \bibinfo{title}{Merger of binary neutron stars to a black hole: disk
  mass, short gamma-ray bursts, and quasinormal mode ringing}.
\newblock \emph{\bibinfo{journal}{Physical Review D - Particles, Fields,
  Gravitation and Cosmology}} \textbf{\bibinfo{volume}{73}},
  \bibinfo{pages}{064027} (\bibinfo{year}{2006}).

\bibitem{rbg+10}
\bibinfo{author}{Rezzolla, L.}, \bibinfo{author}{Baiotti, L.},
  \bibinfo{author}{Giacomazzo, B.}, \bibinfo{author}{Link, D.} \&
  \bibinfo{author}{Font, J.~A.}
\newblock \bibinfo{title}{{Accurate evolutions of unequal-mass neutron-star
  binaries: properties of the torus and short GRB engines}}.
\newblock \emph{\bibinfo{journal}{Classical and Quantum Gravity}}
  \textbf{\bibinfo{volume}{27}}, \bibinfo{pages}{114105}
  (\bibinfo{year}{2010}).

\bibitem{du17}
\bibinfo{author}{Dietrich, T.} \& \bibinfo{author}{Ujevic, M.}
\newblock \bibinfo{title}{{Modeling dynamical ejecta from binary neutron star
  mergers and implications for electromagnetic counterparts}}.
\newblock \emph{\bibinfo{journal}{Classical and Quantum Gravity}}
  \textbf{\bibinfo{volume}{34}}, \bibinfo{pages}{105014}
  (\bibinfo{year}{2017}).

\bibitem{llp+16}
\bibinfo{author}{Lehner, L.} \emph{et~al.}
\newblock \bibinfo{title}{{Unequal mass binary neutron star mergers and
  multimessenger signals}}.
\newblock \emph{\bibinfo{journal}{Classical and Quantum Gravity}}
  \textbf{\bibinfo{volume}{33}}, \bibinfo{pages}{184002}
  (\bibinfo{year}{2016}).

\bibitem{tlf+13}
\bibinfo{author}{{Tanvir}, N.~R.} \emph{et~al.}
\newblock \bibinfo{title}{{A `kilonova' associated with the short-duration
  {\ensuremath{\gamma}}-ray burst GRB 130603B}}.
\newblock \emph{\bibinfo{journal}{Nature}} \textbf{\bibinfo{volume}{500}},
  \bibinfo{pages}{547--549} (\bibinfo{year}{2013}).

\bibitem{jbp+15}
\bibinfo{author}{Just, O.}, \bibinfo{author}{Bauswein, A.},
  \bibinfo{author}{Pulpillo, R.~A.}, \bibinfo{author}{Goriely, S.} \&
  \bibinfo{author}{Janka, H.-T.}
\newblock \bibinfo{title}{{Comprehensive nucleosynthesis analysis for ejecta of
  compact binary mergers}}.
\newblock \emph{\bibinfo{journal}{Monthly Notices of the Royal Astronomical
  Society}} \textbf{\bibinfo{volume}{448}}, \bibinfo{pages}{541--567}
  (\bibinfo{year}{2015}).

\bibitem{rms+09}
\bibinfo{author}{Read, J.~S.} \emph{et~al.}
\newblock \bibinfo{title}{Measuring the neutron star equation of state with
  gravitational wave observations}.
\newblock \emph{\bibinfo{journal}{Phys. Rev. D}} \textbf{\bibinfo{volume}{79}},
  \bibinfo{pages}{124033} (\bibinfo{year}{2009}).

\bibitem{lw15}
\bibinfo{author}{Lackey, B.~D.} \& \bibinfo{author}{Wade, L.}
\newblock \bibinfo{title}{Reconstructing the neutron-star equation of state
  with gravitational-wave detectors from a realistic population of inspiralling
  binary neutron stars}.
\newblock \emph{\bibinfo{journal}{Phys. Rev. D}} \textbf{\bibinfo{volume}{91}},
  \bibinfo{pages}{043002} (\bibinfo{year}{2015}).

\bibitem{amd+15}
\bibinfo{author}{Agathos, M.} \emph{et~al.}
\newblock \bibinfo{title}{Constraining the neutron star equation of state with
  gravitational wave signals from coalescing binary neutron stars}.
\newblock \emph{\bibinfo{journal}{Phys. Rev. D}} \textbf{\bibinfo{volume}{92}},
  \bibinfo{pages}{023012} (\bibinfo{year}{2015}).

\bibitem{ligo18}
\bibinfo{author}{Abbott, B.~P.} \emph{et~al.}
\newblock \bibinfo{title}{{GW170817:} measurements of neutron star radii and
  equation of state}.
\newblock \emph{\bibinfo{journal}{Physical Review Letters}}
  \textbf{\bibinfo{volume}{121}}, \bibinfo{pages}{161101}
  (\bibinfo{year}{2018}).

\bibitem{ligo17d}
\bibinfo{author}{{The LIGO Scientific Collaboration}} \emph{et~al.}
\newblock \bibinfo{title}{{A gravitational-wave standard siren measurement of
  the Hubble constant}}.
\newblock \emph{\bibinfo{journal}{Nature}} \textbf{\bibinfo{volume}{551}},
  \bibinfo{pages}{85--88} (\bibinfo{year}{2017}).

\bibitem{hng+18}
\bibinfo{author}{{Hotokezaka}, K.} \emph{et~al.}
\newblock \bibinfo{title}{{A Hubble constant measurement from superluminal
  motion of the jet in GW170817}}.
\newblock \emph{\bibinfo{journal}{Nature Astronomy}}
  \textbf{\bibinfo{volume}{3}}, \bibinfo{pages}{940--944}
  (\bibinfo{year}{2019}).

\bibitem{planck18}
\bibinfo{author}{{Planck Collaboration}} \emph{et~al.}
\newblock \bibinfo{title}{{Planck 2018 results. VI. Cosmological parameters}}
  (\bibinfo{year}{2018}).
\newblock \eprint{1807.06209}.

\bibitem{rcy+19}
\bibinfo{author}{Riess, A.~G.}, \bibinfo{author}{Casertano, S.},
  \bibinfo{author}{Yuan, W.}, \bibinfo{author}{Macri, L.~M.} \&
  \bibinfo{author}{Scolnic, D.}
\newblock \bibinfo{title}{{Large} {Magellanic} {Cloud} {Cepheid} standards
  provide a 1{\%} foundation for the determination of the {Hubble} constant and
  stronger evidence for physics beyond $\uplambda${CDM}}.
\newblock \emph{\bibinfo{journal}{The Astrophysical Journal}}
  \textbf{\bibinfo{volume}{876}}, \bibinfo{pages}{85} (\bibinfo{year}{2019}).

\end{thebibliography}

\begin{thebibliography}{10}
\setcounter{enumiv}{42}   

\expandafter\ifx\csname url\endcsname\relax
  \def\url#1{\texttt{#1}}\fi
\expandafter\ifx\csname urlprefix\endcsname\relax\def\urlprefix{URL }\fi
\providecommand{\bibinfo}[2]{#2}
\providecommand{\eprint}[2][]{\url{#2}}

\bibitem{tay92}
\bibinfo{author}{Taylor, J.~H.}
\newblock \bibinfo{title}{Pulsar timing and relativistic gravity}.
\newblock \emph{\bibinfo{journal}{\ptrsa}} \textbf{\bibinfo{volume}{341}},
  \bibinfo{pages}{117--134} (\bibinfo{year}{1992}).

\bibitem{psrchive04}
\bibinfo{author}{{Hotan}, A.~W.}, \bibinfo{author}{{van Straten}, W.} \&
  \bibinfo{author}{{Manchester}, R.~N.}
\newblock \bibinfo{title}{{PSRCHIVE} and {PSRFITS:} an open approach to radio
  pulsar data storage and analysis}.
\newblock \emph{\bibinfo{journal}{\pasa}} \textbf{\bibinfo{volume}{21}},
  \bibinfo{pages}{302--309} (\bibinfo{year}{2004}).
\newblock \eprint{astro-ph/0404549}.

\bibitem{tempourl}
\bibinfo{howpublished}{\url{https://github.com/nanograv/tempo}}.

\bibitem{dd85}
\bibinfo{author}{Damour, T.} \& \bibinfo{author}{Deruelle, N.}
\newblock \bibinfo{title}{General relativistic celestial mechanics of binary
  systems. {I.} the {post-Newtonian} motion}.
\newblock \emph{\bibinfo{journal}{Annales de l'I.H.P. Physique th\'eorique}}
  \textbf{\bibinfo{volume}{43}}, \bibinfo{pages}{107--292}
  (\bibinfo{year}{1985}).

\bibitem{dd86}
\bibinfo{author}{Damour, T.} \& \bibinfo{author}{Deruelle, N.}
\newblock \bibinfo{title}{General relativistic celestial mechanics of binary
  systems. {II.} the {post-newtonian} timing formula}.
\newblock \emph{\bibinfo{journal}{Annales de l'I.H.P. Physique th\'eorique}}
  \textbf{\bibinfo{volume}{44}}, \bibinfo{pages}{263--292}
  (\bibinfo{year}{1986}).

\bibitem{dt92}
\bibinfo{author}{Damour, T.} \& \bibinfo{author}{Taylor, J.~H.}
\newblock \bibinfo{title}{Strong-field tests of relativistic gravity and binary
  pulsars}.
\newblock \emph{\bibinfo{journal}{Phys. Rev. D}} \textbf{\bibinfo{volume}{45}},
  \bibinfo{pages}{1840--1868} (\bibinfo{year}{1992}).

\bibitem{shk70}
\bibinfo{author}{{Shklovskii}, I.~S.}
\newblock \bibinfo{title}{Possible causes of the secular increase in pulsar
  periods}.
\newblock \emph{\bibinfo{journal}{Soviet Astronomy}}
  \textbf{\bibinfo{volume}{13}}, \bibinfo{pages}{562} (\bibinfo{year}{1970}).

\bibitem{nt95}
\bibinfo{author}{{Nice}, D.~J.} \& \bibinfo{author}{{Taylor}, J.~H.}
\newblock \bibinfo{title}{{PSR J2019+2425 and PSR J2322+2057 and the proper
  motions of millisecond pulsars}}.
\newblock \emph{\bibinfo{journal}{\apj}} \textbf{\bibinfo{volume}{441}},
  \bibinfo{pages}{429--435} (\bibinfo{year}{1995}).

\bibitem{pbm+19}
\bibinfo{author}{Perera, B. B.~P.} \emph{et~al.}
\newblock \bibinfo{title}{{The dynamics of Galactic centre pulsars:
  constraining pulsar distances and intrinsic spin-down}}.
\newblock \emph{\bibinfo{journal}{Monthly Notices of the Royal Astronomical
  Society}} \textbf{\bibinfo{volume}{487}}, \bibinfo{pages}{1025--1039}
  (\bibinfo{year}{2019}).

\bibitem{ymw17}
\bibinfo{author}{Yao, J.~M.}, \bibinfo{author}{Manchester, R.~N.} \&
  \bibinfo{author}{Wang, N.}
\newblock \bibinfo{title}{A new electron-densiity model for estimation of
  pulsar and frb distances}.
\newblock \emph{\bibinfo{journal}{The Astrophysical Journal}}
  \textbf{\bibinfo{volume}{835}}, \bibinfo{pages}{29} (\bibinfo{year}{2017}).

\bibitem{kkl03}
\bibinfo{author}{{Kim}, C.}, \bibinfo{author}{{Kalogera}, V.} \&
  \bibinfo{author}{{Lorimer}, D.~R.}
\newblock \bibinfo{title}{The probability distribution of binary pulsar
  coalescence rates. i. double neutron star systems in the galactic field}.
\newblock \emph{\bibinfo{journal}{Astrophys. J.}}
  \textbf{\bibinfo{volume}{584}}, \bibinfo{pages}{985--995}
  (\bibinfo{year}{2003}).

\bibitem{kpm15}
\bibinfo{author}{{Kim}, C.}, \bibinfo{author}{{Perera}, B.~B.~P.} \&
  \bibinfo{author}{{McLaughlin}, M.~A.}
\newblock \bibinfo{title}{{Implications of PSR J0737-3039B for the Galactic
  NS-NS binary merger rate}}.
\newblock \emph{\bibinfo{journal}{\mnras}} \textbf{\bibinfo{volume}{448}},
  \bibinfo{pages}{928--938} (\bibinfo{year}{2015}).

\bibitem{pml19}
\bibinfo{author}{{Pol}, N.}, \bibinfo{author}{{McLaughlin}, M.} \&
  \bibinfo{author}{{Lorimer}, D.~R.}
\newblock \bibinfo{title}{Future prospects for ground-based gravitational-wave
  detectors: The galactic double neutron star merger rate revisited}.
\newblock \emph{\bibinfo{journal}{\apj}} \textbf{\bibinfo{volume}{870}},
  \bibinfo{pages}{71} (\bibinfo{year}{2019}).

\bibitem{pml20}
\bibinfo{author}{{Pol}, N.}, \bibinfo{author}{{McLaughlin}, M.} \&
  \bibinfo{author}{{Lorimer}, D.~R.}
\newblock \bibinfo{title}{{An Updated Galactic Double Neutron Star Merger Rate
  Based on Radio Pulsar Populations}}.
\newblock \emph{\bibinfo{journal}{Research Notes of the American Astronomical
  Society}} \textbf{\bibinfo{volume}{4}}, \bibinfo{pages}{22}
  (\bibinfo{year}{2020}).

\bibitem{2013MNRAS.432.1303B}
\bibinfo{author}{{Bagchi}, M.}, \bibinfo{author}{{Lorimer}, D.~R.} \&
  \bibinfo{author}{{Wolfe}, S.}
\newblock \bibinfo{title}{{On the detectability of eccentric binary pulsars}}.
\newblock \emph{\bibinfo{journal}{\mnras}} \textbf{\bibinfo{volume}{432}},
  \bibinfo{pages}{1303--1314} (\bibinfo{year}{2013}).
\newblock \eprint{1302.4914}.

\end{thebibliography}

\begin{table}
\caption{Measured and derived parameters for PSR~J1913+1102.\label{tab:params}}
\begin{center}
\begin{tabular}{lc}
\hline
Parameter name & Value\\
\hline
Reference epoch (MJD)                &  \mysf{57504.0} \\
Observing time span (MJD)            &  \mysf{56072 - 58747}  \\
Number of arrival time measurements  &  \mysf{2541} \\
Solar system ephemeris used          &  \mysf{DE436} \\
Root mean squared timing residual    &  \mysf{56\ \mu s} \\
Reduced $\chi^2$ of timing fit       &  \mysf{1.01} \\
\hline
Right ascension $\alpha$ (J2000)     &  19h\,13m\,29.05365(9) s \\
Declination $\delta$ (J2000)         &  \mysf{11^\circ 02^\prime 05.7045(22)^{\prime\prime}} \\
Proper motion in $\alpha$            &  \mysf{-3.0(5)\ mas\ yr^{-1}} \\
Proper motion in $\delta$            &  \mysf{-8.7(1.0)\ mas\ yr^{-1}} \\
Pulsar spin period, $P$              &  \mysf{27.2850068680286(19)\ ms} \\
Period derivative, $\dot{P}$         &  \mysf{1.5672(7) \times 10^{-19}\ s\,s^{-1}} \\
Dispersion measure, DM               &  \mysf{339.026(3)\ pc\,cm^{-3}} \\
Orbital period, $\dot{P}_{\rm b}$    &  \mysf{0.2062523345(2)\ d} \\
Projected semimajor axis of the pulsar's orbit, $x$ &  \mysf{1.754635(5)\ light\,s} \\
Orbital eccentricity, $e$            &  \mysf{0.089531(2)} \\
Longitude of periastron, $\omega$    &  \mysf{283.7898(19)^\circ} \\
Epoch of periastron passage, $T_0$ (MJD) &  \mysf{57504.5314530(10)}\\
Total system mass, $M$               &  \mysf{2.8887(6)\ M_\odot} \\
Companion mass, $M_c$                &  \mysf{1.27(3)\ M_\odot} \\ 
Rate of periastron advance$^a$, $\dot{\omega}$  &  \mysf{5.6501(7)^\circ\, yr^{-1}} \\
Einstein delay$^a$, $\gamma$         &  \mysf{0.000471(15)\ s} \\
Orbital decay rate$^a$, $\dot{P}_{\rm b}$  &  \mysf{-4.8(3) \times 10^{-13}\ s\, s^{-1}}  \\ 
\hline
Pulsar mass, $M_p$                   & \mysf{1.62(3)} \\
Mass ratio                           & \mysf{0.78(3)} \\
Orbital inclination angle, $i$       & \mysf{55.3^\circ} \\
Dispersion-derived distance$^b$, $d$      & \mysf{7.14\ kpc} \\
Surface magnetic flux density at the poles, $B_0$               & \mysf{2.1 \times 10^9\ G} \\
Characteristic age, $\tau_c$         & \mysf{2.8\ Gyr} \\
Time to coalescence, $T_c$           & \mysf{470(-14,+15)}\ Myr \\
\hline
\end{tabular}
\end{center} 

Values in parentheses represent the $1\sigma$ uncertainty on the last quoted digit. Unless otherwise noted, measured parameters were determined using the Damour \& Deruelle General Relativity (DDGR) timing model\cite{dd85,dd86}, which assumes general relativity to be the correct theory of gravity. \\$^a$Post-Keplerian orbital parameters were measured using the model-independent Damour \& Deruelle (DD) timing model\cite{dd85,dd86,dt92}.
\\$^b$Distance is derived based on a model of the Galactic ionised electron density\cite{ymw17}
\end{table}

\clearpage

\begin{addendum}
 \item The authors wish to thank S.~Nissanke, T.~Tauris, and B.~Metzger for useful and constructive discussions, as well as the anonymous referees for useful comments. 
 The Arecibo Observatory is operated by the University of Central Florida, Ana G.~M\`{e}ndez-Universidad Metropolitana, and Yang Enterprises under a cooperative agreement with the National Science Foundation (NSF; AST-1744119).
 R.D.F.~and P.C.C.F.~acknowledge the support of the PHAROS COST Action (CA16214).
 S.C., J.M.C., F.C., M.A.M, and N.P.~are members of the NANOGrav Physics Frontiers Center, which is supported by the NSF award number PHY-1430284. S.C. and J.M.C. also acknowledge support from the NSF award AAG-1815242.
 V.M.K.~acknowledges support from an NSERC Discovery grant and Herzberg Award, the Canada Research Chairs program, the Canadian Institute for Advanced Research, and FRQ-NT. 
 E.P.~is a Vanier Canada Graduate Scholar. I.H.S.~acknowledges support for pulsar research at the University of British Columbia by an NSERC Discovery Grant and by the Canadian Institute for Advanced Research. J.v.L.~acknowledges funding from Vici research programme `ARGO' (n. 639.043.815) financed by NWO.
 
 \item[Author contributions] R.D.F.~contributed to timing analysis, wrote and ran mass probability contour code, led successful proposals for conducting the observing campaigns for this pulsar with the Arecibo telescope, composed the manuscript, and produced Figures 1 and 3.  P.C.C.F.~led the timing analysis. B.P.P.P.~ and N.P.~led the population synthesis analysis and produced Figure 2.  All authors contributed to discussion regarding the content of this manuscript.
 
 \item[Data and code availability] All data is available from the authors on request. Code used in this analysis is public on Github at the following URLs:
 \begin{itemize}
  \setlength\itemsep{-1px}    
    \item[] \textit{Pulsar timing analysis:} https://github.com/nanograv/tempo
    \item[] \textit{Population synthesis:} https://github.com/NihanPol/2018-DNS-merger-rate
    \item[] \textit{Plotting tools:} https://github.com/rferdman/pypsr
 \end{itemize}
 
 \item[Competing Interests] The authors declare that they have no competing financial interests.

 \item[Correspondence and request for materials] Correspondence and requests for materials
should be addressed to R.D.F.~(email: r.ferdman@uea.ac.uk).
\end{addendum}


\end{document}